\documentclass{article}
\usepackage{spconf,amsmath,graphicx}
\usepackage{booktabs}
\usepackage{multirow}
\usepackage{adjustbox}
\usepackage{float}
\usepackage{subfig}
\usepackage{graphicx}
\newcommand{\squeezeup}{\vspace{-2.5mm}}

\newcommand*{\affaddr}[1]{#1} 
\newcommand*{\affmark}[1][*]{\textsuperscript{#1}}
\title{On the benefit of parameter-driven approaches for the modeling and the prediction of Satisfied User Ratio for compressed video}

\name{Jingwen Zhu\affmark[1], Patrick Le Callet\affmark[1], Anne-Flore Perrin\affmark[2], Sriram Sethuraman\affmark[3], Kumar Rahul\affmark[3]} \vspace{-2.5mm}
\address{\affaddr{\affmark[1]Nantes Université, Ecole Centrale Nantes, CNRS, LS2N, UMR 6004, Nantes, France} \\ \affaddr{\affmark[2]CAPACITÉS SAS, Nantes, France} \  \ \ \affaddr{\affmark[3]Amazon Prime Video, Bangalore, India}  } 
\vspace{-2.5mm}
\begin{document}
%
\maketitle

\begin{abstract}
The human eye cannot perceive small pixel changes in images or videos until a certain threshold of distortion. In the context of video compression, Just Noticeable Difference (JND) is the smallest distortion level from which the human eye can perceive the difference between reference video and the distorted/compressed one. 
Satisfied-User-Ratio (SUR) curve is the  complementary cumulative distribution function of the individual JNDs of a viewer group.
However, most of the previous works predict each point in SUR curve by using features both from source video and from compressed videos with assumption that the group-based JND annotations follow Gaussian distribution, which is neither practical nor accurate. In this work, we firstly compared various common functions for SUR curve modeling. Afterwards, we proposed a novel parameter-driven method to predict the video-wise SUR from video features. Besides, we compared the prediction results of source-only features based (SRC-based) models and source plus compressed videos features (SRC+PVS-based) models.
\end{abstract}
%
\section{Introduction}
\label{sec:intro}

 With the rapid growth of multimedia demand, Picture Wise Just Noticeable Difference (PW-JND) \cite{liu2019deep, tian2020just, shen2020just,lin2020subjective} and Video Wise Just Noticeable Difference (VW-JND) \cite{wang2018prediction, wang2018analysis,zhang2020satisfied,ki2018learning} have been investigated from human perceptive aspects in order to provide high quality of experience for end-users with limited storage and internet bandwidth. Furthermore, JND estimation is helpful for visual processing tasks such as visual signal enhancement and perceptual quality evaluation.

JND depends on 3 factors : (1) display setting, \textit{e.g.,} viewing distance \cite{nakasu1996statistical}, monitor profiling, \textit{etc.}; (2) subjects; (3) image/video contents \cite{lin2015experimental}. In this work, we concentrated on how different video contents impact the location of JND in the context of video compression.
Factor (1) and (2) are fixed and are not investigated.

For a given visual content, JND of different subjects will be different \cite{lin2015experimental}. Wang \textit{et al.}\cite{wang2016mcl} proposed to conduct the subjective test of JND with respect to a viewer group other than the very few experts (golden eyes) for the worst-case analysis, because the group-based quality of experience (QoE) is closer to the realistic applications.
Satisfied-User-Ratio (SUR) curve can be derived from this group-based JND value. SUR curve is defined as the Q-function supposing that the group-based JND follows Gaussian distribution \cite{wang2018prediction}. Intuitively, the value of SUR curve at a certain distortion level $d$, is the percentage of the group users who cannot perceive any difference between the reference stimuli and the distorted stimulus whose distortion level is smaller than $d$, \textit{i.e.,} these users are satisfied.
At a given threshold $p$ for SUR, the corresponding distortion level is defined as $p$\%SUR instead of the misleading notation $p$\%JND in previous works \cite{wang2018prediction,wang2018analysis,zhang2020satisfied,wang2017videoset}.

It is well known that subjective test is expensive and time-consuming, especially for VW-JND. Therefore, it is crucial to develop VW-JND prediction methods in order to predict the encoding parameter (\textit{e.g,} Quantization Parameter (QP)) corresponding to a given SUR threshold. The issue has been addressed in \cite{wang2018prediction}, who 
 proposed a model to predict SUR curve by using support vector regression (SVR) under the assumption that the individual JND points of a group users follow a normal distribution. Their model infers SUR values from VMAF \cite{li2016toward} Quality Degradation Features concatenated with Masking Effect Features \cite{hu2015compressed, hu2016objective} and is trained on the large scale JND dataset of  compressed video \cite{wang2017videoset}.
The 75\%SUR point can then be derived by the predicted SUR curve.
\cite{wang2018analysis} is the extended work of \cite{wang2018prediction}, the 
\textit{2nd} and \textit{3rd} JND points are predicted using 3 different settings in which the reference inputs of the predictor are different.
Instead of predicting encoding parameter QP as SUR profile, Zhang \textit{et al.} \cite{zhang2020satisfied} proposed a novel perceptual model to predict SUR versus bitrate, which is more widely used in practice. Three kinds of features, Masking features, re-compression features and basic attribute features, are extracted from original reference video to build a feature vector, which will be used to conduct a Gaussian Processes Regression (GPR) to predict SUR. 

However, previous works assume that the individual VW-JND of a group viewers follows Gaussian distribution, which might not be the optimal modeling method. Besides, when predicting SUR curve, previous works are computationally expensive because they extract features from SRC and from every encoded PVS and predict the individual SUR score of each PVS to derive the SUR curve. Therefore, we first investigate the modeling of SUR curve and then propose a novel SUR prediction model only based on SRC.  
\squeezeup
\section{Modeling \& prediction of SUR}
\label{sec:format} 
\begin{figure}[t]
	\centering
	\includegraphics[width=1.0\columnwidth]{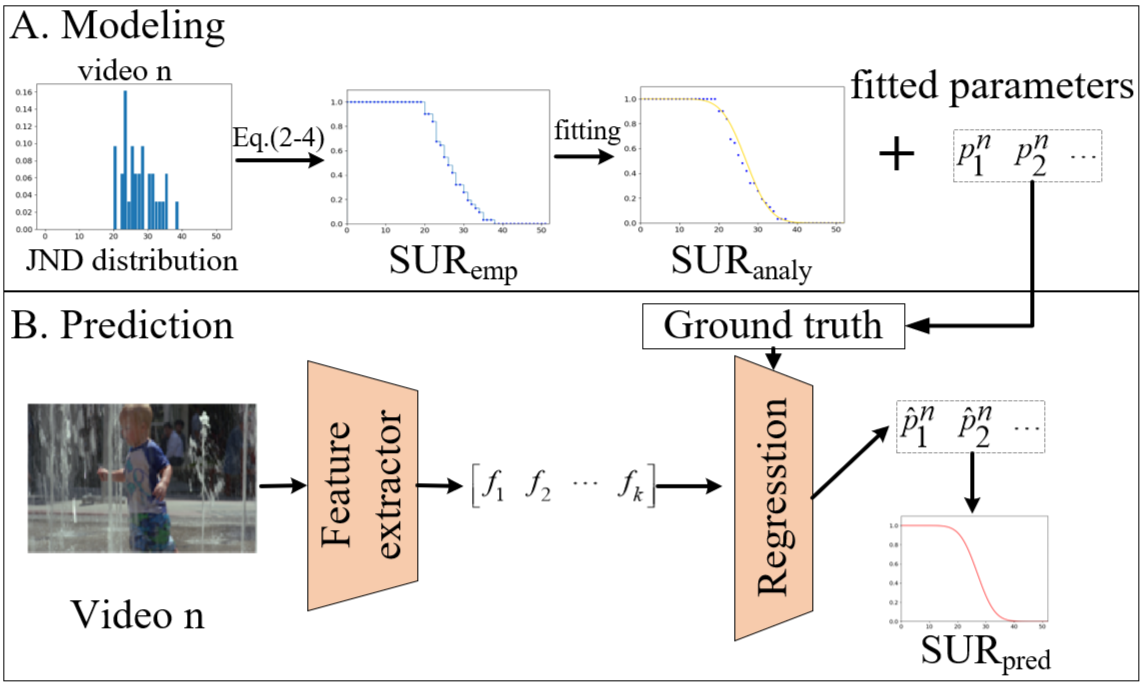}   
	\caption{Illustration of the pipeline of SUR and JND modeling (A) and prediction (B)}
	\vspace{-5.0mm}
	\label{fig:fm}
\end{figure}

In this work, we firstly investigated the modeling of the group-based VW-JND rather than using a simple normality test as in the previous works \cite{wang2017videoset, wang2016mcl}, in order to find the mathematical model that best fits SUR. Afterwards, we proposed a SUR prediction method via predicting the model parameters obtained by the modeling fit, 
which is called parameter-driven model, in preference to the commonly used point-by-point models. The entire pipeline is shown in Fig.\ref{fig:fm}

\vspace{-2.5mm}
\subsection{Modeling of SUR}
\label{sec:modeling}
\begin{figure}[htb]
	\centering
	\includegraphics[width=1.0\columnwidth]{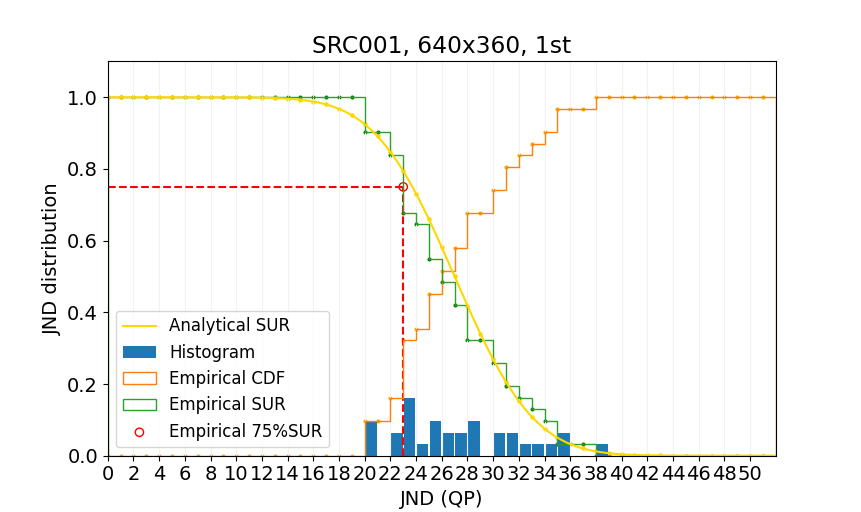} 
	\caption{Distribution of group-based VW-JND (blue bar); empirical and analytical SUR (in green and yellow respectively) and empirical 75\%SUR (red circle) of SRC001 (360p) in VideoSet \cite{wang2017videoset}} 
	\vspace{-5.0mm}
	\label{fig:sur_demo}
\end{figure}

When modeling the SUR of VW-JND, previous works \cite{wang2017videoset,wang2016mcl} conducted Jarque-Beta test \cite{jarque1987test} to verify the normality of the VW-JND position of every subject. However, when revisiting the original distribution of VW-JND annotations, we found that Gaussian distribution is not necessarily the best modeling of the VW-JND distribution  (Fig.\ref{fig:sur_demo}). Therefore, we first clarify the definition of empirical SUR curve and p\%SUR. Afterwards, we set the Complementary Cumulative Distribution Function (CCDF) of different candidate distributions (\textit{e.g.,} Gaussian, Sigmoid, Weilbull, \textit{etc.}) as model functions to find the best fit function of the empirical SUR curve. The model functions were named as "analytical SUR".
\setlength{\belowdisplayskip}{4pt} \setlength{\belowdisplayshortskip}{4pt}
\setlength{\abovedisplayskip}{4pt} \setlength{\abovedisplayshortskip}{4pt}
For a given content $m$, assuming that there are $N$ reliable subjects' VW-JND annotation, VW-JND of a subject "n" is denoted by $j_n^m$. VW-JND of $N$ subjects can be denoted by $J^m$ as
\begin{equation} \label{eq:samples}
   {J^m} = \left[ {j_1^m,j_2^m,...,j_N^m} \right].
\end{equation}
Considering $J^m$ as a discrete random variable,
the Probability Mass Function (PMF) of $J^m$ is defined by
\begin{equation} \label{eq:distribution}
p(x) = P({\rm{JND = x}}) = \frac{1}{N}\sum\limits_{i = 1}^N {{\boldsymbol{1}}\left( {{j_i} = x} \right)},
\end{equation}
where
$\boldsymbol{1}(c)$ is an indicator function that equals to 1 if the specified binary clause $c$ is true. Thus, the empirical Cumulative Distribution Function (CDF) can be calculated from the PMF:
\begin{equation} \label{eq:cdf_emp}
{\rm{CDF}_{emp}}(x) = P(\rm{JND} \le x) = \sum\limits_{\omega  < x} {p(\omega )}.
\end{equation}
Because $J$ is discrete, the CDF increases only by jump discontinuities, as shown in Fig.\ref{fig:sur_demo} (orange curve).
The empirical SUR (green curve in Fig.\ref{fig:sur_demo}) is the CCDF of $J$:
\begin{equation} \label{eq:sur_emp}
{\rm{SUR}_{emp}}(x) = 1 - \rm{CDF}_{emp}(x).
\end{equation}
The empirical $p$\%SUR is defined as:
\begin{equation}\label{eq:jnd_emp}
p\% {\rm{SUR}_{{\rm{emp}}}} = \min \left\{ {x\left |  {{\rm{SUR}_{{\rm{emp}}}}\left( x \right)} \right. \le p\% } \right\},
\end{equation}
where the range of x is the range of distortion levels. For instance, the range of x is [0, 51] with step equals to 1 in VideoSet.
%
The analytical SUR curve and p\%SUR are calculated by Eq.(\ref{eq:sur_analy}) and (\ref{eq:jnd_ana}), $f(x)$ is the Probability Density Function (PDF). Contrary to the empirical SUR, the analytical SUR is a continuous function.
\begin{equation}  \label{eq:sur_analy}
    {\rm{SUR}_{{\rm{analy}}}}\left( x \right) = 1 - {\rm{CDF}_{{\rm{analy}}}}(x) = 1 - \int\limits_{ - \infty }^\infty  {f(x)dx}. 
\end{equation}
\squeezeup
\begin{equation} \label{eq:jnd_ana}
    {p\% \rm{SUR}_{{\rm{analy}}}} = \mathop {\arg \min }\limits_{x \in \{ 0,1,...51\} } ({\rm{SUR}_{{\rm{analy}}}}(x) - p\% )
\end{equation}
After computing the empirical SUR from the VW-JND annotations, we fitted the discrete points in empirical SUR with 8 model functions (Table\ref{tab:model_functions}) for each video. It can be easily proved that the CDF (Eq.(\ref{eq:cdf_emp})) of a distribution is monotonic non-decreasing, thus SUR (Eq.(\ref{eq:sur_emp})) is monotonic non-increasing. Therefore, monotonic constraint was applied during least-squares optimization for polynomial model function.
\begin{table}[t]
\begin{center} 
\small
\setlength{\intextsep}{0pt}
\caption{Summary of candidate model functions. (NB para is the number of parameters in model function)} \label{tab:model_functions}
\begin{tabular}{c|c|c}
\hline
Name         & Model function    & NB para\\ \hline
Polynomial-3 & \multirow{2}{*}{$f(x) = \sum\limits_{k = 0}^n {{a_k}{x^k}}$} & 4             \\ \cline{1-1} \cline{3-3} 
Polynomial-4 &                   & 5             \\ \hline
Gaussian     &$1 - \frac{1}{2}\left( {1 + erf\left( {\frac{{x - \mu }}{{\sigma \sqrt 2 }}} \right)} \right)$                   & 2             \\ \hline
2-para-logistic &$1 - \frac{1}{{1 + {e^{ - (x - \mu )/s}}}}$                   & 2             \\ \hline
4-para-logistic &  $f\left( x \right) = b + \frac{L}{{1 + {e^{ - k(x - {x_0})}}}}$                 & 4             \\ \hline
Weibull      & ${e^{ - {{\left( {\frac{x}{\lambda }} \right)}^k}}}$                  & 2             \\ \hline
Gumbel       &  $1 - {e^{ - {e^{ - (x - \mu )/\beta }}}}$                 & 2             \\ \hline
Rayleigh     & ${e^{\frac{{ - {x^2}}}{{\left( {2{\sigma ^2}} \right)}}}}$                  & 1             \\ \hline
\end{tabular}
\vspace{-5.0mm}
\end{center}
\end{table}
In addition to MAE and RMSE, we use $\Delta p\% {\rm{SUR}_{\left| {E - A} \right|}}$ (Eq.(\ref{eq:delta_JND_EA})) to evaluate the candidate model functions, where $p\%$ is set to 75\%. The experiment results will be detailed in Section \ref{sec:exp.1}.
\begin{equation}  \label{eq:delta_JND_EA}
    \Delta p\% {\rm{SUR}_{\left| {E - A} \right|}} = \left| {p\% {\rm{SUR}_{{\rm{emp}}}} - p\% {\rm{SUR}_{{\rm{analy}}}}} \right|
\end{equation}
\vspace*{-2.5mm}\squeezeup
\subsection{Prediction of SUR}
\label{sec:pred}
We revisited the SUR prediction model proposed by Wang \textit{et al.} \cite{wang2018prediction} namely the baseline. Furthermore, we analysed the two main drawbacks of the baseline model and proposed solutions to solve these issues. 

\begin{figure}[t]
	\centering
	\includegraphics[width=1.0\columnwidth]{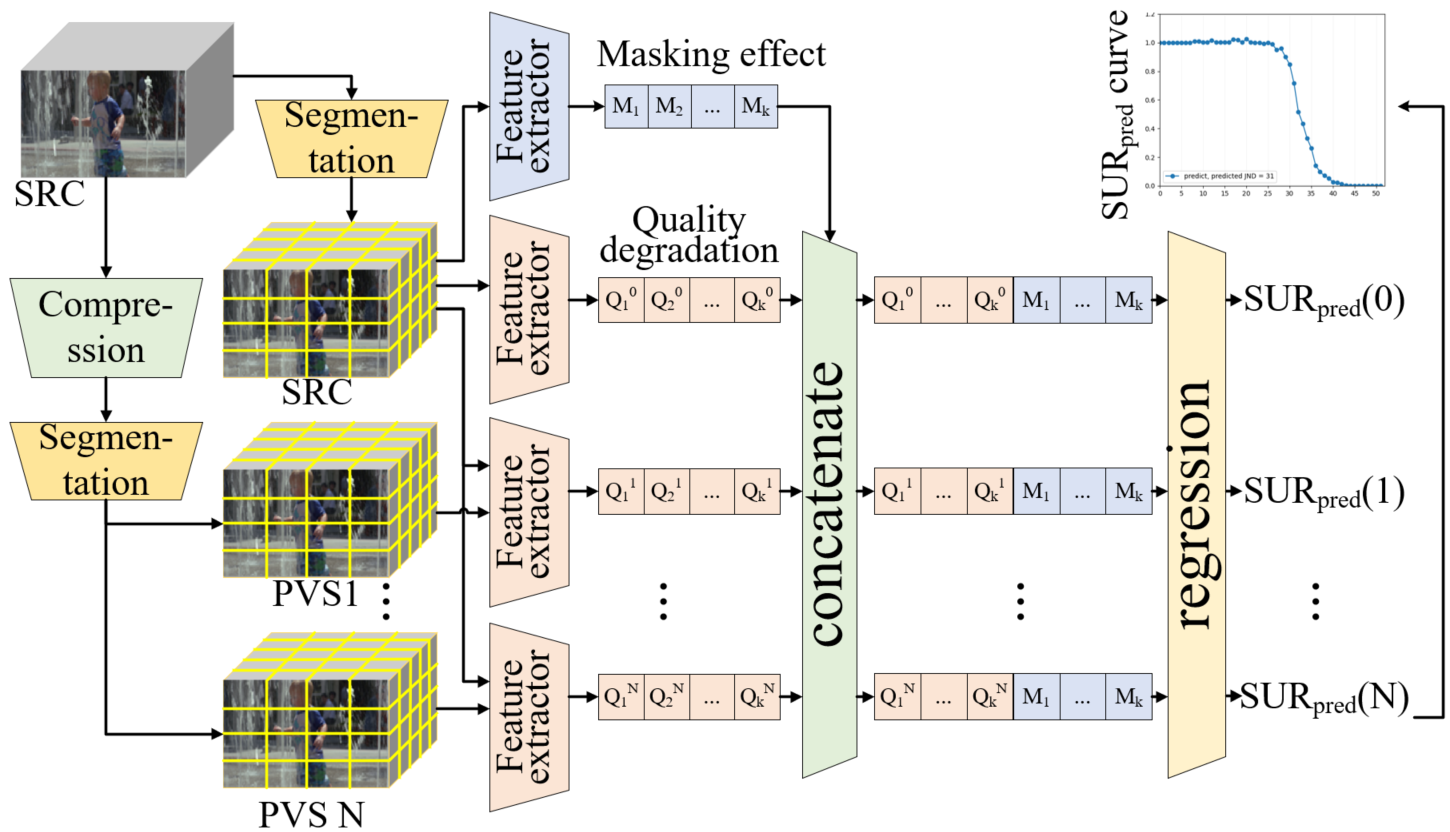}   
	\caption{Illustration of baseline SUR prediction model}
	\label{fig:baseline}
	\vspace{-4.0mm}
\end{figure}

As shown in Fig.(\ref{fig:baseline}), the input of the baseline model is the uncompressed source video (SRC). SRC is firstly compressed with different encoding parameters (\textit{e.g.},QP 1 to 51) to get a series of PVS (Processed Video Sequence). Afterwards, SRC and all PVS are segmented to small video patches both spatially and temporally to extract features from the eye fixation level. Two types of features are extracted from the segmented patches: Masking effect and Quality degradation ($M$ and $Q$ in Fig.(\ref{fig:baseline})). Masking effect is a measure of the spatial and temporal randomness \cite{hu2015compressed, hu2016objective}. A high randomness masks distortions for human eye, \textit{i.e.}, it indicates the human eye hardly perceive the difference. Quality degradation is calculated based on the difference of quality scores (\textit{e.g.}, VMAF) between SRC and PVS. The masking effect and quality degradation feature vectors of one video are the histogram and cumulative curve of its video patches, respectively. When extracting the Quality degradation features, only video patches with significant quality degradation were selected to compute the final feature vector. The two feature vectors for SRC and each PVS are concatenated and are used to predict SUR scores by regression.

The baseline model is computational expensive as individual SUR score of each PVS of a SRC must be computed to derive the SUR curve prediction. Accordingly, features from a SRC and its PVSs (\textit{i.e.,} 52 sequences) have to be computed.
This is the first drawback of the baseline model.
The second drawback is that the $\rm{SUR_{pred}}$ curve of baseline model is not monotonic non-increasing because every individual point of SUR curve is predicted separately.
The basic meaning of SUR is: for a given distortion level $x$, its SUR value is what percentage of subjects are satisfied, \textit{i.e.,} what percentage of subjects do not perceive the difference between the reference video and all the distorted video whose distortion level is less than $x$. Therefore, it is not reasonable that when the distortion level increases, the SUR value increases.

To address these issues, we proposed a straightforward solution with the help of the modeling parameters of SUR in Section \ref{sec:modeling}.
The pipeline of the proposed method is shown in Fig.\ref{fig:fm}. Instead of predicting SUR scores of every PVS and geting the SUR curve accordingly, the parameters which describe the SUR curve (e.g., $\sigma$ and  $\mu$ for Gaussian) are predicted. Only SRC is used for prediction, hence these models are called \textbf{SRC-based} model. Masking effect features are extracted from SRC and Support Vector Regression (SVR) \cite{smola2004tutorial} is used for regression. Although SRC-based models are preferred in real-life applications, it is still interesting to understand how important is the quality degradation information from PVSs to the prediction of SUR and JND. Therefore, we also investigated \textbf{SRC+PVS-based} model, where masking effect and quality degradation features of every PVS are concatenated into one vector for regression to predict the modeling parameters.  

\section{Experiments and results}
\label{sec:exp}
\subsection{Modeling}
\label{sec:exp.1}
In our experiment, VideoSet \cite{wang2017videoset} was used to evaluate the modeling and prediction of SUR and JND. VideoSet include 220 5-second SRCs in 4 resolutions. Each SRC is encoded with H.264 codec with QP from 1 to 51. More than 30 subjects participated into the viewer group for each SRC and each individual VW-JND annotation was publicly available as well. As mentioned in Section \ref{sec:modeling}, we used the individual VW-JND annotations of each SRC to generate the $\rm{SUR}_{{\rm{emp}}}$ and $\rm{75\%SUR}_{{\rm{emp}}}$ (Eq.(\ref{eq:distribution})-(\ref{eq:jnd_emp})) and every discrete points of $\rm{SUR}_{{\rm{emp}}}$ were used to fit the model functions listed in Table \ref{tab:model_functions}. Scikit-learn linear regression was used for polynomial fittings and monotonic constraints were applied by using Polyfit\footnote{https://github.com/dschmitz89/Polyfit}. Non-linear least squares from SciPy were used for other model functions.
\begin{table}[htb]
\caption{Mean of MAE, RMSE and $\Delta {\rm{75\% SUR}_{\left| {E - A} \right|}}$ for different model functions with VideoSet \cite{wang2017videoset}}
\begin{adjustbox}{width=200pt,center}
\centering
\label{tab:modeling}
\begin{tabular}{c|c|c|c} 
\hline
Name            & MAE             & RMSE            & $\Delta {\rm{75\%SUR}_{\left| {E - A} \right|}}$               \\ 
\hline
Polynomial-3    & 0.1204          & 0.1466          & 5.0614           \\
Polynomial-4    & 0.1085          & 0.1338          & 4.7420           \\
Gaussian        & 0.0147          & 0.0253          & 0.6625           \\
2-para-logistic & 0.0156          & 0.0250          & 0.5875           \\
4-para-logistic & 0.0164          & \textbf{0.0236} & \textbf{0.5761}  \\
Weibull         & \textbf{0.0138} & 0.0240          & 0.6761           \\
Gumbel          & 0.0220          & 0.0343          & 0.5977           \\
Rayleigh        & 0.1451          & 0.1703          & 8.9114           \\
\hline
\end{tabular}
\end{adjustbox}
\vspace{-4.0mm}
\end{table}

For every SRC in 4 resolutions of VideoSet ($220 \times 4=880$ SRC in total), we calculated the MAE and RMSE between $\rm{SUR}_{\rm{emp}}$ and $\rm{SUR}_{\rm{analy}}$ and also the difference between empirical and analytical 75\%SUR : $\Delta {\rm{75\%SUR}_{\left| {E - A} \right|}}$ (Eq.(\ref{eq:delta_JND_EA})) with different fitting functions. The results are shown in Table \ref{tab:modeling}. It can be observed that the CCDF of Gaussian distribution is not the best modeling for SUR. 4-para-logistic model function outperforms the other candidate model functions both in RMSE and $\Delta {\rm{75\%SUR}_{\left| {E - A} \right|}}$. 

\subsection{Prediction}
\begin{figure}[htb]
	\centering
	\includegraphics[width=1.0\columnwidth]{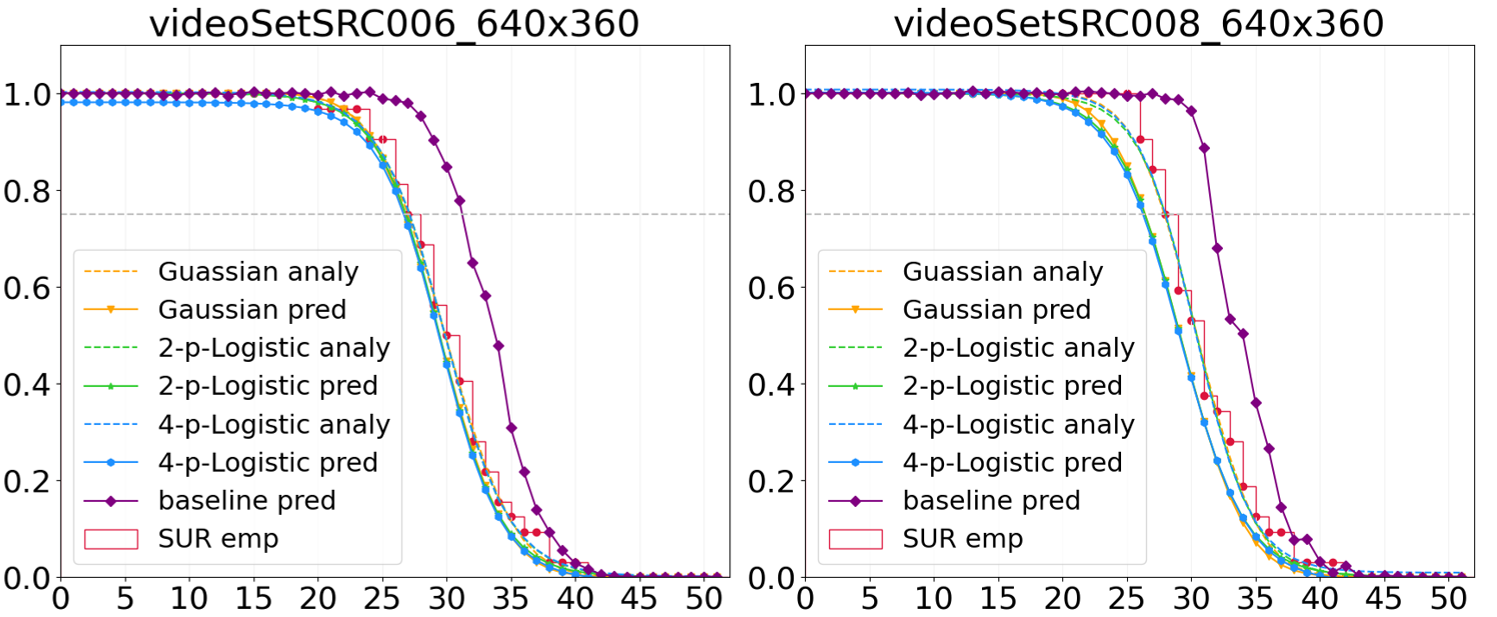}   
	\caption{Examples of SUR prediction results comparison between SRC-based models and baseline model}
	\label{fig:visual}
\end{figure}

Each prediction model was evaluated on videos using 5-fold cross validation. Radial basis function kernel was used for SVR. We firstly compare the baseline model (in-house implementation) and 3 SRC-based parameter-driven models with different model functions. Fig.\ref{fig:visual} shows the SUR prediction results of 2 SRCs. The dashed lines in orange, green and blue are the analytical SUR curves obtained from fitting to the red points in empirical SUR with Gaussian, 2-p-logistic and 4-p-logistic respectively. Plain lines with dots are the predicted SUR of the 3 SRC-based models obtained by predicting the parameters of the corresponding dashed lines. The purple line is the SUR prediction of baseline model. 
Difference between \textbf{P}redicted and \textbf{A}nalytical SUR (denoted $\Delta {\rm{SUR}_{\left| {{\rm{P - A}}} \right|}}$) is the MAE between them. $\Delta {\rm{SUR}_{\left| {{\rm{P - A}}} \right|}}$ indicates the error between ground truth (analytical SUR curve) and prediction SUR curve, but the modeling error (between empirical and analytical SUR curve) are not considered. Therefore, difference between \textbf{P}redicted and \textbf{E}mpirical SUR (denoted $\Delta {\rm{SUR}_{\left| {{\rm{P - E}}} \right|}}$) is evaluated as well. $\Delta {\rm{75\%SUR}}$ is evaluated in the same way. The results are shown in Table \ref{tab:res_1}. The 3 SRC-based parameter-driven models outperform the baseline model both in $\Delta {\rm{SUR}}$ and $\Delta {\rm{75\%SUR}}$. The prediction errors between Gaussian and logistic parameter-driven models are quite close. However, the 4-para-logistic which has the smallest modeling error performs worse than Gaussian and 2-para-logistic. 

\begin{table}[htb]
\caption{Averaged prediction error comparison between baseline model and 3 SRC-based parameter-driven models.}
\label{tab:res_1}
\begin{adjustbox}{width=\columnwidth,center}
\centering
\begin{tabular}{l|l|l|l|l|l} 
\hline
\multirow{2}{*}{RES}   & \multirow{2}{*}{Model
  name} & \multicolumn{2}{c|}{$\Delta {\rm{SUR}}$}            & \multicolumn{2}{c}{$\Delta {\rm{75\%SUR}}$}         \\ 
\cline{3-6}
                       &                               & $\left| {{\rm{P-A}}} \right|$ & $\left| {{\rm{P-E}}} \right|$ & $\left| {{\rm{P-A}}} \right|$ & $\left| {{\rm{P-E}}} \right|$  \\ 
\hline
\multirow{4}{*}{360p}  & baseline                      & 0.0769                   & 0.0799                   & 4.3682                   & 4.3773                    \\
                       & 2-p-Gaussian                  & \textbf{0.0459}          & \textbf{0.0480}          & 2.4773\textbf{}          & 2.5864                    \\
                       & 2-p-Logistic                  & 0.0462                   & 0.0489\textbf{}          & \textbf{2.4455}          & \textbf{2.5682}           \\
                       & 4-p-Logistic                  & 0.0496                   & 0.0515                   & 2.4591                   & 2.5909                    \\ 
\hline
\multirow{4}{*}{540p}  & baseline                      & 0.0786                   & 0.0812                   & 4.3182                   & 4.2909                    \\
                       & 2-p-Gaussian               & \textbf{0.0397}          & \textbf{0.0428}          & 2.1182\textbf{}          & 2.1045                    \\
                       & 2-p-Logistic               & 0.0398                   & 0.0437\textbf{}          & \textbf{1.9727}          & \textbf{2.0955}           \\
                       & 4-p-Logistic               & 0.0435                   & 0.0458                   & 2.0045                   & 2.1000                    \\ 
\hline
\multirow{4}{*}{720p}  & baseline                      & 0.0783                   & 0.0820                   & 4.2864                   & 4.2909                    \\
                       & 2-p-Gaussian               & \textbf{0.0433}          & \textbf{0.0447}          & \textbf{2.1636}          & \textbf{2.2045}           \\
                       & 2-p-Logistic               & 0.0435                   & 0.0459\textbf{}          & \textbf{2.1636}          & 2.2364                    \\
                       & 4-p-Logistic               & 0.0467                   & 0.0476\textbf{}          & \textbf{2.1636}          & 2.2318                    \\ 
\hline
\multirow{4}{*}{1080p} & baseline                      & 0.0801                   & 0.0834                   & 4.6000                   & 4.5591                    \\
                       & 2-p-Gaussian               & 0.0412                   & \textbf{0.0431}          & 2.3455\textbf{}          & 2.2136                    \\
                       & 2-p-Logistic               & \textbf{0.0409}          & 0.0440\textbf{}          & \textbf{2.1182}          & \textbf{2.1773}           \\
                       & 4-p-Logistic               & 0.0439                   & 0.0455                   & 2.1455                   & 2.1727                    \\
\hline
\end{tabular}
\end{adjustbox}
\end{table}

We also compared SRC-based model and SRC+PVS-based model, the results are shown in Table \ref{tab:res_2}. It can be observed that adding quality degradation information from PVSs will improve the prediction of both SUR and 75\%SUR, but with a cost of encoding SRC to 51 PVSs.  
\begin{table}[htb]
\caption{Averaged prediction error comparison between SRC-based and SRC+PVS based model on 1080p with Guassian modeling.}
\label{tab:res_2}
\begin{adjustbox}{width=230pt,center}
\centering
\vspace{-2.5mm}
\begin{tabular}{c|c|c|c|c} 
\hline
\multirow{2}{*}{Model} & \multicolumn{2}{c|}{$\Delta {\rm{SUR}}$}                      & \multicolumn{2}{c}{$\Delta {\rm{75\%SUR}}$}                   \\ 
\cline{2-5}
                       & $\left| {{\rm{P-A}}} \right|$ & $\left| {{\rm{P-E}}} \right|$ & $\left| {{\rm{P-A}}} \right|$ & $\left| {{\rm{P-E}}} \right|$  \\ 
\hline
SRC-based              & 0.0412                        & 0.0431                        & 2.3455                        & 2.2136                         \\
SRC+PVS-based         & \textbf{0.0377}               & \textbf{0.0412}               & \textbf{2.0727}               & \textbf{2.1409}                \\
\hline
\end{tabular}
\end{adjustbox}
\end{table}

\section{Conclusion}
\label{sec:conclu}
In this paper, we proposed a novel pipeline for SUR modeling and prediction to predict the optimal encoding parameter only from SRC. Experiment results show that the proposed parameter-driven model (2-p-Logistic for instance) improves the mean SUR prediction error to 0.046, reducing it by 43.64\% compared with the baseline and reduces the mean 75\%SUR prediction error from 4.38 QP (baseline) to 2.27 QP. 
Furthermore, compared with SRC-based model, the SRC+PVS-based model slightly improves the mean prediction error of SUR curve and 75\%SUR by 0.0019 and 0.0727 QP respectively, which means the quality degradation features from PVSs are not crucial to SUR prediction.
\vfill\pagebreak

\newpage
\bibliographystyle{IEEEbib}
\bibliography{refs}

\end{document}